# Media Content Atlas

A Pipeline to Explore and Investigate Multidimensional Media Space using Multimodal LLMs

mediacontentatlas.github.io

Merve Cerit, Eric Zelikman, Mu-Jung Cho, Thomas N. Robinson, Byron Reeves, Nilam Ram, Nick Haber

Stanford University

As digital media use continues to evolve and influence various aspects of life, developing flexible and scalable tools to study complex media experiences is essential. This study introduces the Media Content Atlas (MCA), a novel pipeline designed to help researchers investigate large-scale screen data beyond traditional screen-use metrics. Leveraging multimodal large language models (MLLMs), MCA enables moment-by-moment content analysis, content-based clustering, topic modeling, image retrieval, and interactive visualizations. Evaluated on 1.12 million smartphone screenshots continuously captured during screen use from 112 adults over an entire month, MCA facilitates open-ended exploration and hypothesis generation as well as hypothesis-driven investigations at an unprecedented scale. Expert evaluators underscored its usability and potential for research and intervention design, with clustering results rated 96% relevant and descriptions 83% accurate. By bridging methodological possibilities with domain-specific needs, MCA accelerates both inductive and deductive inquiry, presenting new opportunities for media and HCI research.

## 1 INTRODUCTION

As individuals spend an average of over 8 hours a day on screens [68], digital media mirrors and molds individual behaviors, decisions, and cultural norms [56]. Despite the heterogeneity of media experiences, most research relies on aggregate metrics (e.g., screen time on a particular app), overlooking the multidimensional nature of media content [24,53]. Reliance on simplistic definitions of media experience has likely hindered the development of comprehensive theories and targeted interventions [15]. Advances in screen data collection now enable continuous observation of everything that appears on users' screens as they go about their daily lives (e.g., [52,70]), offering unparalleled opportunities for understanding human behaviors on and beyond screens.

However, analyzing such data is challenging due to its highly private nature, sheer volume, and the multimodal complexity of text, images, videos, and interface design elements. Existing approaches, including text-based analysis using optical character recognition (OCR) [7,8,64] or metadata extraction [2,49] and qualitative methods that require manual labeling [28,29,45], are labor-intensive and impractical for such multimodal datasets. Hence, while many relevant datasets have been collected, most studies are limited to analyses that only explore a fraction of the media captured.

To address these challenges, we propose leveraging recent advancements in open-source multimodal large language models (MLLMs) [73] that integrate text and image analysis within a unified framework. MLLMs enable integration of image description, clustering, topic extraction, visual question answering, zero-shot classification, and natural language-based image retrieval and search into customized domain-specific applications that can maintain the privacy of sensitive data. Given these capabilities, we developed a pipeline to support researchers in conducting exploratory analysis, generating inductive hypotheses, and creating visualizations—all without the need for extensive manual labeling.

In this study, we make the following key contributions:

- We introduce the Media Content Atlas (MCA)—a customizable pipeline powered by open-source multimodal models (e.g., CLIP, LLaVA-One Vision). MCA facilitates content-based clustering, topic modeling, image retrieval, and interactive visualization, enabling open-ended exploration and inductive inquiry. We demonstrate MCA's utility with 1.12 million in situ screens from 112 participants, uncovering insights that go far beyond traditional media content categories.

- We present initial evaluation results from domain experts in media-related fields, who assessed the pipeline's usability, utility, and accuracy through think-aloud sessions and surveys. They found MCA to be highly usable, more informative, and descriptive than traditional methods, very likely to be adopted into their workflows, and very likely to support the discovery of new media content patterns. They rated clustering and topic labeling as 96% relevant, and description accuracy as 83% accurate.
- We outline opportunities and considerations to refine and integrate the pipeline into media research and beyond, including design and implementation suggestions for hypothesis generation, interventions, and deeper exploration of large-scale multimodal data.

## 2 BACKGROUND & RELATED WORK

***Observing media behavior at scale.*** Studying digital media behaviors requires moving beyond self-reports and retrospective assessments to real-time capture of content in real-life contexts [15,54]. Researchers explored methods to observe digital interactions at a granular level—ranging from screen time [18,47] to keystroke data [25,32,69]. In recent years, continuous screen capture methods that unobtrusively record smartphone screenshots during active use [10,53,70] have emerged as a novel life-logging [17] and digital phenotyping method [21]. The resulting datasets provide large-scale, detailed observations of how individuals move through, interact with, and engage with media content over time.

***Organizing and describing the content in multidimensional media space.*** Describing digital media content is challenging due to its context-dependent, multimodal, and ever-changing nature; "Any media experience, even a short one, is infinitely describable" [53,55], largely because much of real life (e.g., relationships, health, entertainment) is now digital. Prior research has explored media content through various lenses, such as problems addressed [53], sentiment [27], activity [42,65,71], and visual complexity [9,20]. In addition, commercial content classifications (e.g., by Google Play Market labels) provide predefined app names and app categories (e.g., Social, Communication), but they often fail to capture the full range of affordances and behaviors reflected in content [4]. Researchers have used qualitative content analysis [1,29], OCR [7,8,64] with sentiment analysis [74], and topic modeling [36,46] to support both inductive research—by iteratively identifying themes—and deductive inquiry, applying predefined criteria to detect specific content types [23,49,57], including suicidal thinking [22], emotional valence and informational content [9], and payday loan advertisements [30].

***Use of multimodal large language models for content understanding.*** Recent advances in open-source multimodal large language models (MLLMs) have transformed the analysis of diverse image types, shortcutting the need for extensive manual content labeling. CLIP [51], a self-supervised model trained on 400 million image-text pairs from the Internet without human annotation, enables robust multimodal understanding across diverse domains. Similarly, LLaVa-One Vision [33], designed as a vision-and-language assistant, achieved state-of-the-art performance in visual question-answering. MLLMs have been widely applied in content understanding tasks, including biomedical [34] and pathology image analysis [12], content moderation [72], and cyberbullying detection [58]. Additionally, LLM agents capable of understanding web and smartphone UIs to accomplish tasks are an emerging research area [6,26]. Despite their potential, MLLMs have yet to be utilized for analyzing large-scale smartphone screen data—rich in diverse content and UI elements.

***Designing interactive systems for data exploration and inductive inquiry.*** Inductive inquiry in media content analysis requires tools that support open-ended exploration and match the enormous complexity of the experiences the tools must represent. Prior research in network systems [19,62] and bioinformatics [44,50] has shown that interactive visualizations of large-scale, complex datasets [13] can reveal unexpected patterns and generate novel insights. Extending from the work in these fields, our work aims to design and build an interactive pipeline that, following best design practices [61], facilitates researchers' discovery process while securely handling the private screen data from which the discoveries are derived. Invoking Shneiderman's framework for creativity [60], we designed an interactive system that might allow users to experience "aha" moments [59] by approaching problems from multiple perspectives, zooming in and out of the data, and by offering low floors and high ceiling [48] simultaneously providing beginners with accessibility and experts with advanced functionalities and customization.

## 3 METHODS

***Smartphone Screenshot Data.*** We processed and built the pipeline on 1.12 million smartphone screenshots and log data from 112 participants (10,000 screenshots each; demographics in Appendix) collected over one month (December 15, 2020 – January 15, 2021). This data is a subset of a larger data repository collected at the Stanford Human Screenome Project [54]. After providing informed consent, participants installed software that unobtrusively captured screenshots every 5 seconds when the screen was active, alongside foreground app metadata. These multimodal data were encrypted, compressed, and securely transmitted to a research server, following IRB-approved privacy and security protocols. Participants were compensated $15 for each two weeks of data collection.

***Media Content Atlas Pipeline.*** The MCA pipeline follows four key steps: (1) embedding and description generation, converting complex images into numerical representations and text descriptions; (2) clustering and topic modeling to organize content in interpretable topics; (3) image retrieval for semantic search; and (4) interactive visualization for exploration. The pipeline is model-agnostic and can be customized through further prompting or fine-tuning. Our code is available here.

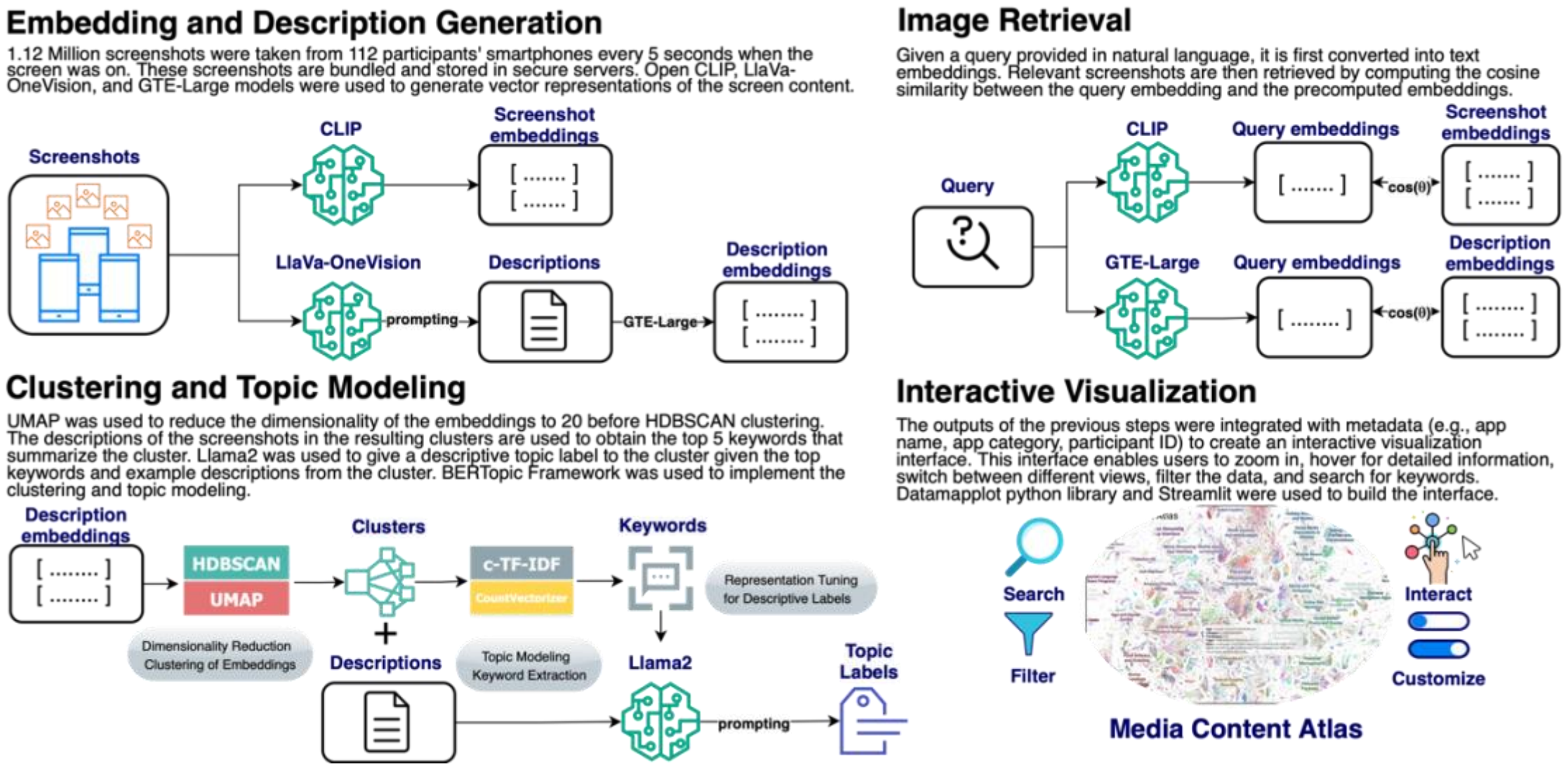

Figure 1: Media Content Atlas Pipeline Architecture Diagram (more information in Appendix).

***Preliminary Tests and Evaluation: Testing with Experts.*** To evaluate and iterate on our design, we structured our tests around the following key questions:

- What types of clusters does this pipeline generate, and how do they vary across dimensions (e.g., app-based, UI elements (e.g., layouts), or higher-order themes (e.g., political content)?
- How does this pipeline compare to typical clustering methods, such as grouping by app names (e.g., Instagram, Google Chrome) or app categories (e.g., social, gaming)?
- How relevant are the topic names generated by the pipeline to the images they represent?
- How accurate is Llava-OneVision in generating content descriptions for screen data?
- How similar are the images within each cluster in terms of visual or content characteristics?
- How well do the models retrieve relevant images for a wide range of queries, from concrete (e.g., food, blood) to abstract (e.g., gambling, depression) concepts?
- What use cases could this pipeline support to advance media research and deepen our understanding of human behaviors on and beyond screens?

To address these through exploratory and inductive inquiry in real-world research while safeguarding highly sensitive data, we recruited four experts (3rd – 6th coauthors) with IRB-approved access. Their expertise spans media processes and effects (coauthors 3–5), media psychology (coauthors 4 & 5), medicine (coauthor 3), developmental psychology (coauthors 5 & 6), computer science (coauthor 6), and education (coauthors 4 & 6). Experts participated in think-aloud sessions with interactive visualizations and completed three surveys: **(S1) Exploration & Comparison with Baseline**, assessing informativeness, descriptiveness, and usefulness of MCA compared to baseline (app-based); **(S2) Clustering & Topic Modeling**, evaluating topic-image relevancy, description accuracy, and within-cluster similarity; and **(S3) Image Retrieval & Applications**, rating retrieval quality and discussing potential use cases. Clusters and images used in the surveys were randomly selected but remained consistent across experts to ensure comparability. Three surveys (~1 hour each survey), accessed via a secure local web app, were conducted in person or over Zoom. The first author observed experts' interactions with the tools, obtained their think-aloud, and provided technical support as needed.

## 4 RESULTS

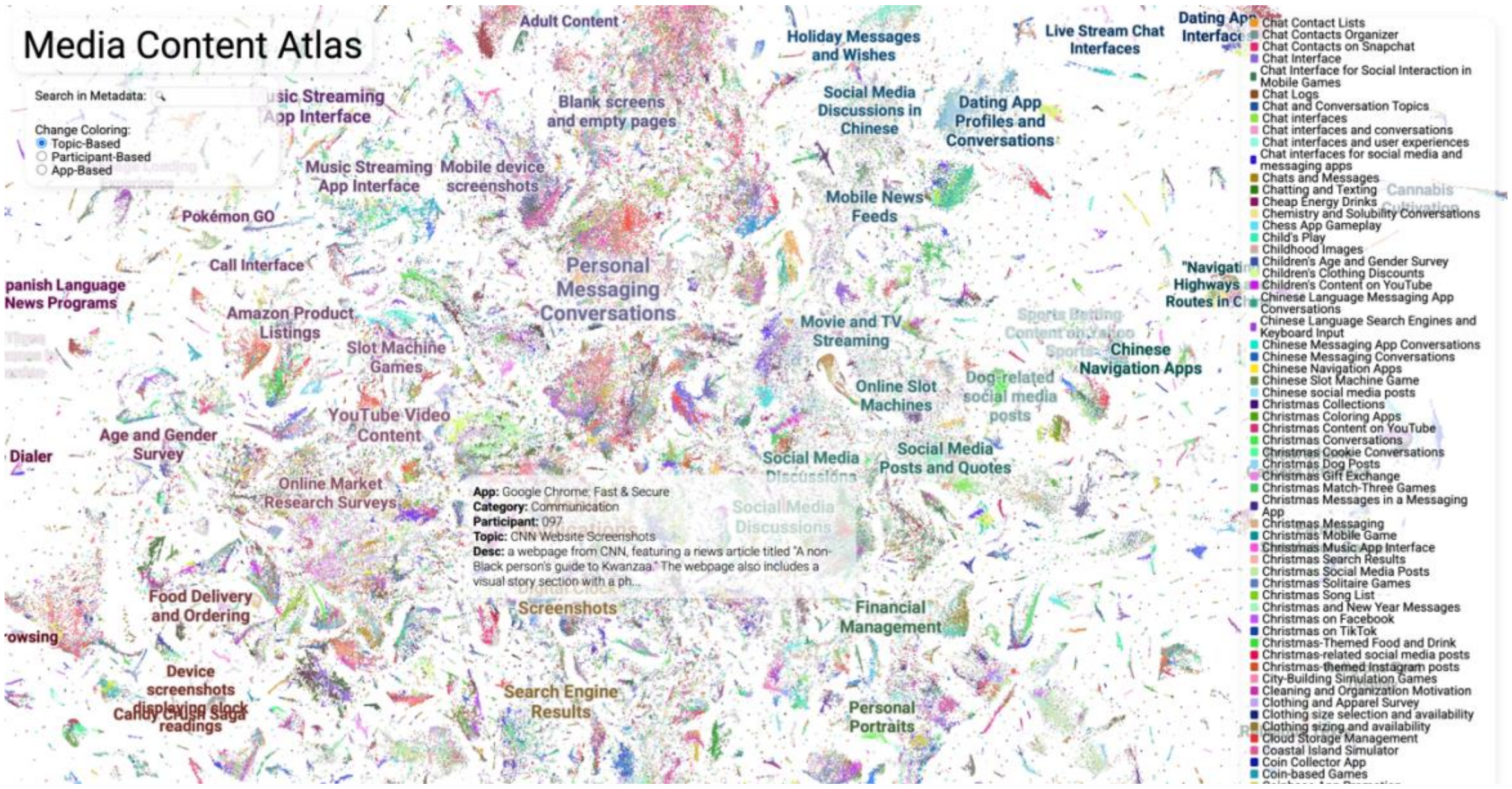

Figure 2: A screenshot of the interactive Media Content Atlas dashboard interface.

### 4.1 Overview of Clusters and Topic Labels

Media Content Atlas generated (under default settings) 2,086 clusters from 1.12 million screenshot images, with 232,375 images (20.75%) classified as noise—data points in low-density regions that do not belong to any significant cluster. Cluster sizes varied significantly, from 24 screenshots in smaller themes such as "Fishing Games" to 20,018 in broader categories like "Personal Messaging Conversations," with an average of 426 screenshots per cluster (SD = 1,906). In answering Q1, we found that the clusters captured diverse behavioral patterns and content types.

Some clusters were app-specific, such as "Candy Crush Saga," while others distinguished content within apps, including "TikTok Kitchen Content." Others described content spanning multiple platforms, addressing themes like "ASMR Content on Social Media" and "Dogs Social Media Posts," or identified UI elements, such as "Phone Keypad Interface." Some clusters were focused on content surrounding specific events, such as "Capitol Riots and Protests (January 2021)," and "COVID-19 Information," while others were focused on specific interests, including "Adult Content Streaming," "Sports Betting," "Math and Science Communication", "Kantian Philosophy." (more examples in Appendix) Altogether, the generated clusters confirmed and highlighted the extreme complexity and multidimensionality of the media space.

The resulting clusters and topic labels from the pipeline are visualized in an interactive interface (Figure 2), built using DataMapPlot [38]. Each point represents a screenshot, colored by its topic (this can be changed using the controls in the top left). Users can zoom, drag, search, and filter metadata keywords and switch views by topic, participant, app, or app category. Larger clusters appear initially, revealing finer clusters upon zooming. An alphabetically ordered legend allows filtering by checking the corresponding boxes. Hovering over a point reveals detailed metadata, including the app, app category, participant ID, topic, and a short description (in Figure 2 by the text box in the lower middle part of the image).

### 4.2 Preliminary Tests and Evaluation: Expert Surveys

***Exploration of the Media Content Atlas and Comparison with the Baseline.*** To address Q2, experts explored the baseline view of the interactive visualization, where data was colored by app names and app categories, and then the content-based MCA view (Figure 2), where data was colored by generated topics. They were then asked to compare the two views in terms of informativeness, descriptiveness, and usefulness for their research, using a 7-point Likert scale where 1 indicated a stronger preference for the baseline and 7 indicated a stronger preference for the content-based MCA clustering. Experts found the content-based clustering to be more informative (all responses 6 and 7), more descriptive (all responses 6 and 7), and more useful for their research (all responses 6 and 7). When asked about adoption likelihood, all experts indicated that they would be "very likely" to use the MCA.

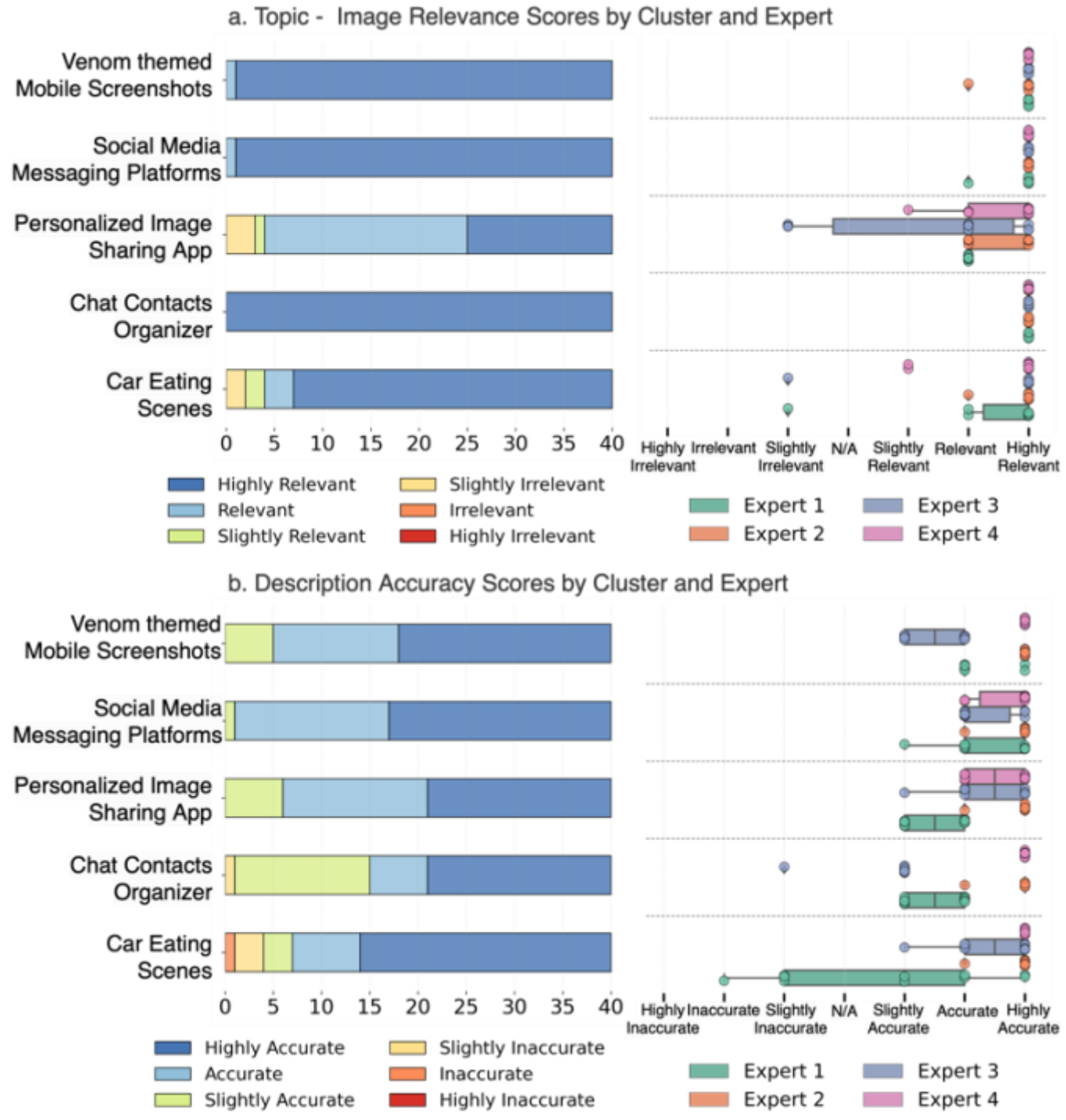

Figure 3: **Topic-Image Relevance(a) and Description Accuracy(b) Scores by Cluster and Expert.** The bar plots on the left represent the count of responses for each randomly selected cluster, with varying bar lengths indicating the number of responses and colors representing different slider values. The boxplots with scatter points on the right illustrate the distribution of scores provided by each of the four experts.

***Topic-Image Relevancy and Description Accuracy.*** To address Q3 and Q4, experts reviewed five randomly selected clusters with their topic labels. Each cluster included ten randomly selected images and their corresponding descriptions. Experts rated the relevance of the images to the topic labels and the accuracy of the descriptions using a 7-point Likert scale presented as a slider. For the pipeline-produced topic labels, **96%** were rated as **"Relevant"** or **"Highly Relevant."** For the text descriptions generated by LlaVA-OneVision, **83%** were rated as **"Accurate"** or **"Highly Accurate."**

While the experts generally agreed on the relevance of clusters, different categories of content posed some challenges. In the *Personalized Image Sharing App* cluster, while experts found the screenshots broadly aligned with the concept, the overlap with other categories such as video calls and social media posts, led to variability in relevance ratings. The model's inability to distinguish between image-based content and video calls based on UI elements contributed to lower accuracy scores. Similarly, the *Car Eating Scenes* cluster presented challenges in identifying the act of eating, with some images containing food but lacking explicit eating cues. Experts also noted instances where the model's descriptions included "hallucinated" elements, such as references to steering wheels and dashboards that were not actually in the images, as well as misinterpretations of perspective (see specific examples in the Appendix).

***Within-Cluster Content Similarity.*** To address Q5, we randomly selected five clusters. For each cluster, experts were shown 25 randomly selected images, and an additional image from the same cluster was displayed to the left. Without any descriptions or topic names, they were asked: "How similar do you think the content of the screenshot on the left is to the group of screenshots shown on the right?" The results indicated that **89%** of the images were rated as **"Similar"** or **"Highly Similar"** to the other images in their respective clusters.

Clusters with highly similar content, such as Cluster 2 (black lock screens with a lock icon and timestamps), had minimal variance, while more diverse clusters, like Cluster 3 (a mix of wallpapers and text messages), led to greater variance. Experts used various criteria for similarity evaluation, including themes, colors, objects, visual features, text, and UI elements. For example, Cluster 1 (loading or blank screens) was assessed based on the overarching theme of 'lack of content,' while Cluster 4 (health-related messaging UI) was evaluated by both UI and message content. In Cluster 5 (car maintenance images), experts varied in their focus—some prioritized thematic coherence, while others prioritized visual elements like scene composition and human presence.

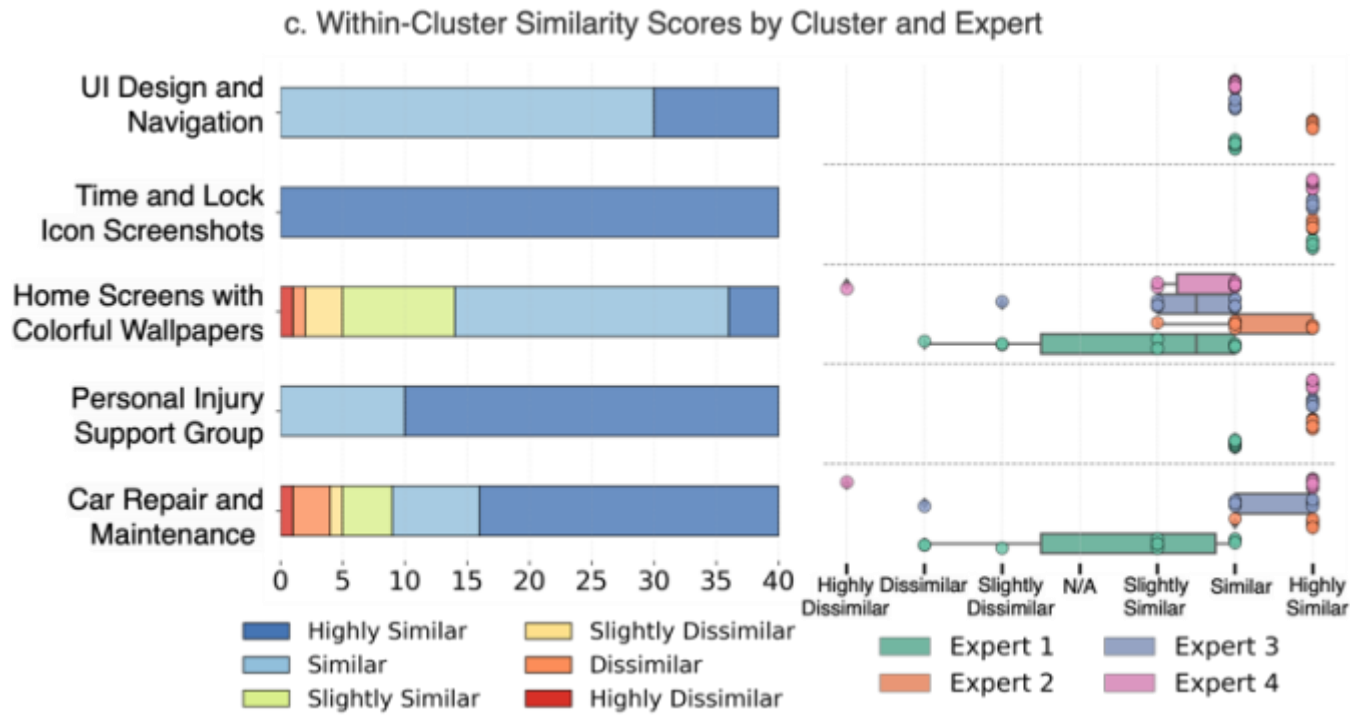

Figure 4: **Within-Cluster Similarity by Cluster and Expert.** The bar plots on the left represent the count of responses for each randomly selected cluster, with varying bar lengths indicating the number of responses and colors representing different slider values. The boxplots with scatter points on the right illustrate the distribution of scores provided by each of the four experts for images within the five cluster topics.

***Image Retrieval based on Predetermined Queries.*** To address Q6, experts were presented with 10 images per query. Each set included the top 5 images retrieved by CLIP and 5 by Llava-OneVision+GTE-Large, with no indication of source. Experts were asked to rate the relevance of each image to the given query (e.g., "An image with {query} content"). Overall, **79.5%** of images were rated as either **Highly Relevant (73%)** or **Relevant (6.5%)**. Specifically, 80.5% of images retrieved by CLIP and 78.5% of images retrieved by Llava-OneVision+GTE-Large were rated as either Highly Relevant or Relevant.

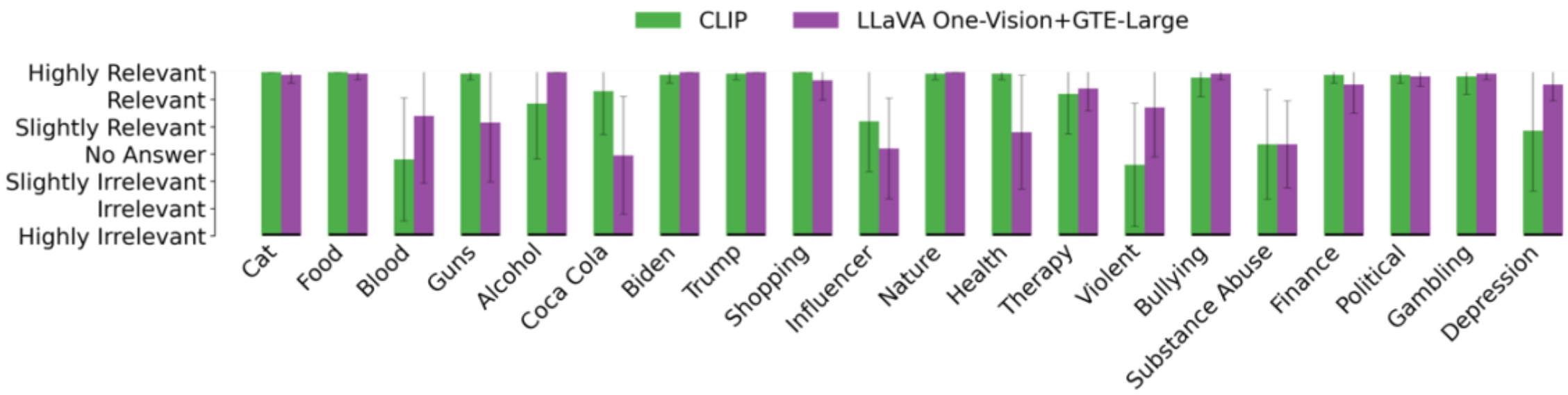

Figure 5: Mean expert ratings for each query, comparing CLIP (green bars) and Llava-OneVision+GTE-Large (purple bars). Higher bars indicate higher relevance scores. Queries are arranged from concrete (e.g., cat) to abstract (e.g., depression).

Both models performed well across various queries, retrieving relevant images for concrete and abstract concepts. Queries like Cat, Food, and Political yielded accurate results, capturing diverse textual, visual, and symbolic elements. However, some queries posed challenges—CLIP often misclassified ketchup as blood, causing false positives, while Llava-OneVision+GTE-large lacked specificity in the Coca-Cola query, retrieving general beverage-related images instead of brand-specific ones (more in Appendix). Queries on mental health constructs, such as Depression and Substance Abuse, illustrated their complexity [3,11,14,31,41], with model outputs reflecting common symptoms and associated terminology. Similarly, the Influencer query often returned ads or promotional content rather than influencer-specific material, suggesting the topic/query may be under-defined. These findings provide insights into the models' underlying training data and embedding mechanisms, highlighting the need for precise queries, clearer construct delineations, and refined prompts to improve accuracy.

***Potential Use Cases and Expert Insights.*** In response to Q7, experts identified several applications of the Media Content Atlas (MCA), highlighting potential use in descriptive analysis, comparative studies, and intervention evaluation. They emphasized the need to explore how users navigate within the media space, with one noting, "We can start by examining how various kinds of content appear—this is not yet known." MCA facilitates both individual and thematic analyses, enabling the identification of content patterns (e.g., climate change, political ads). The pipeline supports both between- and within-person analyses, providing insights into individual and group-level variations over time. One expert posed, "How can we compare and calculate a correlation between an individual's space with aggregate spaces: volume and spread of the personal content space, over time?" MCA can aid hypothesis generation and intervention design by mapping content themes with clinical outcome data. Experts noted its potential to identify causal relationships and design interventions, with one suggesting, "We can hypothesize potentially causal associations and intervention strategies between observed patterns and outcomes of interest, such as health and mental health, then design and test those interventions."; and its potential for supporting rapid comparisons between control and intervention groups.

## 5 DISCUSSION AND KEY LESSONS

**The Media Content Atlas enabled the discovery of previously unknown and unexpected content characteristics from 1.12 million smartphone screens—an amount that would require more than 190 continuous 8-hour workdays for a researcher to simply view at 5 seconds per screenshot.** This significantly expands the possibilities for describing and analyzing media content at scale. By offering a flexible and customizable toolset, MCA enables researchers to explore media beyond conventional methods. One expert described this exploratory process as "inductive play," highlighting the iterative and hands-on nature of our design, which fosters familiarity with the data and encourages experimentation.

**Testing MCA provided experts with a clearer understanding of the data and the models' capabilities and limitations, enabling them to better articulate design needs and identify analytical gaps.** Structured questions from a nonexpert may have further facilitated this process by encouraging collaborative reflection, ensuring researchers examined the data through a shared lens. Some experts prioritized higher-order themes, while others concentrated on specific features, including colors, objects present, and human representation. One expert emphasized the importance of refining descriptions of food and drinks, while another focused on detailed representation of age, race, and gender. One noted that viewing additional images could refine the studied constructs. These insights demonstrate the iterative nature of our approach, where inductive inquiry refines focus and expectations as understanding of the phenomenon evolves.

**Our testing with experts revealed several critical design principles for tools supporting large-scale screen data analysis, emphasizing secure data handling, user-friendly exploration, customizability, seamless dissemination, scalability, and interoperability.** Ensuring protected environments for analysis, providing intuitive interfaces, and allowing researchers to tailor models and visualizations to their research needs are key factors. Seamless dissemination enables stakeholders to share insights without compromising privacy, while scalability ensures efficient processing of vast datasets, though further testing is needed with larger data volumes. Standardized formats for prompts, metadata schemas, and code structures enhance interoperability and facilitate collaboration by enabling consistent data analysis. These principles will guide future developments to address the complexities of multimodal screen data analysis at scale.

## 6 LIMITATIONS AND FUTURE WORK

Key limitations include constraints related to underlying models [66], hyperparameters, and prompt design, which impact results. Future work should explore alternative models and refined prompts to better meet research needs. Expert evaluation bias is another limitation, as evaluators are also co-authors due to data privacy. Future efforts could build representative synthetic benchmark datasets. MCA can support a semi-supervised approach where human validation of LLM-generated clusters refines ground-truth data and identifies edge cases, reducing manual effort. Future work will implement additional modular views (e.g., outcome-based, longitudinal) to further enhance the pipeline's applicability across research contexts.

## 7 CONCLUSION

As media use evolves, developing flexible, scalable catalogues of media experiences is crucial to capturing the diversity of screens people encounter and the multiple ways they can be interpreted. This study introduced Media Content Atlas (MCA), a pipeline for analyzing large volumes of high-complexity screen data. MCA enables researchers to move beyond traditional screen-use metrics and gain deeper insights into digital media complexities through multimodal LLMs. To our knowledge, MCA is the first pipeline to: (1) analyze moment-by-moment screen content, (2) conduct content-based clustering, (3) perform topic modeling, (4) enable image retrieval from theoretical queries, and (5) provide interactive visualizations for multi-perspective exploration. By bridging methodological possibilities with domain-specific needs, MCA accelerates both inductive and deductive inquiry, offering broader implications for media and HCI research.

## ACKNOWLEDGMENTS

Research reported in this publication was supported in part by the National Heart, Lung, and Blood Institute of the National Institutes of Health under Award Number R01HL169601, and National Science Foundation Award Number 2302701. The content is solely the responsibility of the authors and does not necessarily represent the official views of the National Institutes of Health. This research was also supported by the Stanford University Department of Pediatrics, the Stanford Maternal and Child Health Research Institute, the Precision Health and Integrated Diagnostics (PHIND) Center at Stanford, a Stanford RISE COVID-19 Crisis Response Faculty Seed Grant, MediaX at Stanford, the McCoy Family Center for Ethics in Society, the Ethics, Society and Technology Hub, the Stanford Institute for Human-Centered Artificial Intelligence, and the Stanford Graduate School of Education Transforming Learning Accelerator. The first author is funded by the Stanford Graduate Interdisciplinary Fellowship.

# A APPENDICES

## Smartphone Screen Data - Participant Demographics

The dataset consists of 1,120,000 screenshots that were contributed by 112 participants with an average age of 47.1 years (SD = 15.55), ranging from 20 to 78 years. The gender distribution indicates that most participants are female (54.5%), followed by male (43.8%), and a small percentage identifying as other or preferring not to answer (1.7%). The majority of participants identified as White (69.6%), reflecting a substantial portion of the sample. Other racial/ethnic groups included African American/Black (8.9%), Native American (6.3%), and Asian (5.4%), while Hispanic/Latino participants accounted for 4.5% of the sample. A smaller percentage, 5.3%, identified as belonging to multiple racial/ethnic groups or selected "Other." Regarding marital status, 47.3% of participants reported being married or living as married, while single/never married individuals accounted for 28.6%, and divorced or separated participants made up 19.6%. A small proportion, 4.5%, reported being widowed. In terms of education, the most common level attained was a 4-year college degree (approximately 29.5%), followed by those with a master's degree (23.2%), and some college education (16.1%). Participants with a doctoral or professional degree represented a smaller proportion (around 8.9% and 5.4%, respectively). Income levels were diverse, with the most common income bracket being $15,000 - $24,999, representing 20.5% of respondents. Additionally, 16.1% reported earning $50,000 - $74,999, while 13.4% fell into the highest income bracket of $100,000 or more. Notably, 7.1% preferred not to disclose their income. Regionally, participants were distributed across the U.S., with the South being the most represented region, followed by the Midwest, the West, and the Northeast.

## Pipeline Development Process and Implementation Considerations

### *Embedding and Description Generation*

**CLIP** (Contrastive Language-Image Pre-training) [51], specifically clip-ViT-L-14, was used to generate embeddings (vector representations) of the screenshots. CLIP is a self-supervised multimodal model that is trained on 400M image-text pairs from the Internet, trained without human annotation. The model maps text and images to a shared vector space and can perform a wide variety of zero-shot retrieval use cases.

**Llava-OneVision** [33], an open-source large multimodal model that is built to be a large vision-and-language assistant (LlaVA) [35] with high visual question-answering capabilities across single-image, multi-image, and video scenarios. Specifically, the llava-onevision-qwen2-7b-ov-hf model is used to generate an image description of each screenshot, with the following hyperparameters: a max new token of 128 and temperature of 0. The batch inference was done with a batch size of 8, in 10 A100 40 GB GPUs used, it took ~3 seconds per image.

Image description prompt: `"Describe this screenshot in detail, using this output format: The screenshot displays [the content goes here, including verbatim text if present, or a specific description of what the text is about; if there is an image, describe exactly what it depicts]. Focus on explaining the exact content of any text or image. Include the app name if identifiable, but DO NOT include quantitative details, such as likes or status bar information."`

Different prompts were explored, with "Describe this screenshot in detail" yielding the highest variance in output quality, as expected. Some outputs exhibited generalized or subjective comments such as "looking casual and happy," and paraphrased textual content rather than verbatim inclusion. While the model successfully identified some applications just based on UI, such as TikTok and Twitter, it frequently misidentified others if there is no explicit reference to the app name. Additionally, the model tended to overemphasize numerical elements (e.g., likes, comments) and status bar details (e.g., battery level, clock), necessitating explicit prompt refinements to exclude such content. Future improvements include customization of the prompt and increasing the maximum number of output tokens to generate more comprehensive descriptions.

The resulting descriptions were embedded using the open **GTE-large** model (specifically gte-large-en-v1.5) [37], an English text embedding model known for its efficiency and strong benchmark performance. GTE-large has been extensively trained on a large-scale corpus spanning diverse domains and scenarios, making it highly suitable for downstream tasks such as semantic textual similarity and information retrieval. At the time of analysis, GTE-large ranked as the top-performing model under 500M parameters in the MTEB (Massive Text Embedding Benchmark) [43], which was a key factor in its selection for this study.

***Clustering and Topic Modeling***

**BERTopic** [16] , a neural network-based modular framework, was applied for topic modeling. **UMAP** (Uniform Manifold Approximation and Projection for Dimension Reduction) [40], a nonlinear dimensionality reduction technique that protects both local and global structures in the data, is used to reduce the dimensionality of the embeddings before clustering using **HDBSCAN** (Hierarchical Density-Based Spatial Clustering of Applications with Noise) [5,39] HDBSCAN is especially useful with complex noisy data that has different shapes and densities. The dimensionality reduction and clustering process involved the use of UMAP and HDBSCAN. The UMAP hyperparameters were set as follows: n_components was initially set to 20 and subsequently reduced to 10 for clustering, n_neighbors was set to 50 to balance specificity and cluster granularity, and min_dist was set to 0.05 to ensure tighter cluster formation. Additionally, the negative_sample_rate was set to 20, and n_epochs was configured to 1000 to improve convergence at the expense of computational resources. Cosine similarity was chosen as the metric to better capture high-dimensional relationships.

For the clustering phase, HDBSCAN was used with min_cluster_size set to 400 to ensure medium-sized cluster formation, and min_samples set to 20 to define the minimum number of samples required to consider a point core. The clustering process was further refined using the Excess of Mass (EOM) cluster selection method, which optimally selects clusters based on density distribution. The hyperparameters were fine-tuned through an iterative process that incorporated both unsupervised validity index evaluations [39] and qualitative visual assessments. The selection of hyperparameters considered the trade-off between granularity and interpretability of clusters. While smaller clusters provide more granular insights, they can overwhelm users with an excessive number of topics and lead to performance issues in rendering visualizations. Therefore, the optimization process balanced the need for detailed information with the practical constraints of usability and computational efficiency.

**Llama2** [67], specifically Llama-2-7b, was employed for the task of cluster topic labeling due to its balance of efficiency and effectiveness in handling text summarization tasks. The process involved extracting representative documents from each cluster, which were then combined with the top five keywords identified by BERTopic. The model was prompted to generate concise and meaningful labels that accurately captured the thematic essence of each cluster. Given that the model's context length is limited to 4096 tokens, not all descriptions within a big cluster could be accommodated; therefore, a subset of representative descriptions was selected. The selection of representative descriptions is significantly influenced by HDBSCAN's clustering output; if a cluster lacks density or contains high noise, the generated label may not accurately reflect the intended theme. The topic labeling process required balancing cluster granularity, as excessively fine-grained clusters resulted in fragmented topics, while broader clusters led to overly generalized and less informative labels. Future enhancements include investigating methods to merge topic labels based on semantic similarity, integrating predefined taxonomies, and refining prompt engineering to enhance output consistency. Additionally, expanding the context length could improve labeling accuracy. During experimentation with few-shot prompting, it was observed that increasing the number of representative documents occasionally caused the model to output the example label, such as "environmental impacts of eating meat," suggesting that the extended context may disproportionately weigh the example prompt relative to individual documents.

```
system_prompt = """
You are a helpful, respectful, and honest assistant for labeling topics.
<</SYS>>
"""
example_prompt = """
I have a topic that contains the following documents:
- The screenshot displays a website where it says traditional diets in most cultures were primarily plant-based with a little meat on top, but with the rise of industrial style meat production and factory farming, meat has become a staple food.
- The screenshot features an image of meat and the captions read: Meat, but especially beef, is the word food in terms of emissions.
- The screenshot shows a quote on Twitter: "Eating meat doesn't make you a bad person, not eating meat doesn't make you a good one".
The topic is described by the following keywords: 'meat, beef, eating, emissions, food, health, image, processed, climate change, Twitter quote'.

Based on the information about the topic above, please create a short and specific label of this topic, in English. Make sure you only return the label and nothing more.
```

```
[/INST] Environmental impacts of eating meat
"""
main_prompt = """
[INST]
I have a topic that contains the following documents:
[DOCUMENTS]
The topic is described by the following keywords: '[KEYWORDS]'.
Based on the information about the topic above, please create a short and specific label of this topic, in English. Make sure you only return the label and nothing more.
[/INST]
"""
```

***Image Retrieval***

For retrieval, the input query is first embedded to compute the cosine similarity with the embeddings generated from two sources: the original screenshot embeddings obtained directly from the CLIP model and the image description embeddings generated through the Llava-OneVision model in combination with GTE-large. The same query is processed through both embedding methods, and cosine similarity scores are calculated between the query embedding and the corresponding embeddings from the two sources. Based on these similarity scores, the top "k" most similar filenames are returned for each query.

***Interactive Visualization***

The resulting clusters, along with topic labels, were combined with each image's embeddings (reduced to 2D), descriptions, and related smartphone metadata—including foreground app package information, which was mapped to app names and categories using the Google Play Market categorization API. These data were utilized to create interactive visualizations using the DataMapPlot library [38]. The visualizations enable users to search for specific keywords within the metadata, filter results based on apps, app categories, participants, and topics. The library natively supports most of these features and offers flexibility for customization through the addition of HTML, JavaScript, and CSS modules. We customized the library to incorporate additional views tailored to our analysis needs. t-SNE employed dimensionality reduction for visualization to provide a clearer separation of clusters, enhancing interpretability.

**Testing with Experts**

The study employs locally hosted open-source web applications [63] for each survey, enabling expert interaction easily within secure virtual servers to ensure the privacy of screenshot data and facilitate data collection. Throughout the survey, responses are automatically saved to prevent data loss. Experts provide their feedback through a combination of quantitative and qualitative measures, including slider-based questions and open-ended responses. Our code for these surveys can be found here.

***Think aloud & Survey 1 - Exploration & Comparison with Baseline***

Experts begin the survey by learning about the task of comparing typical content classification with the proposed content-based clustering. They are selected for their expertise in media processes and receive guidance on using the tool, encouraged to approach it from their research perspective while thinking aloud. An interactive tutorial with synthetic data demonstrates key features such as hovering, toggling topics, zooming, and searching metadata like app names or keywords.

On the second page, experts explore the baseline visualization of 1.2 million screenshots (shown as scatter points) colored by app name and category. They can switch between visualization modes, including app-based, category-based, and participant-based coloring. The third page presents the content-based clustering view on the same data, grouping screenshots by inferred topics.

The final page gathers expert evaluations of the content-based clustering method through slider-based responses to the following questions: "How informative do you find content-based clustering compared to the baseline (app name and app category)?", "How descriptive is content-based clustering compared to the baseline (app name and app category)?", and "How useful is content-based clustering for your research compared to the baseline (app name and app category)?". Responses range from "Much Less" to "Much More" informative, descriptive, and useful. Experts also assess the likelihood of adopting the method by responding to: "How likely are you to use this method for research in analyzing screen content?", with options from "Very Unlikely" to "Very Likely."

Qualitative feedback is collected through the following open-ended questions: “Any additional comments comparing the two methods?”, “How would this method be useful for your research? Please describe specific scenarios or use cases where this method would be beneficial.”, and “What are your suggestions for the applications of this method? Please provide any suggestions you have for improving the application of this method, including any features or functionalities you would like to see added and any potential challenges or limitations you foresee.”

***Survey 2 - Clustering & Topic Modeling***

**Topic-Image Relevancy and Description Accuracy.** Experts are presented with randomly selected clusters of screenshots that have been grouped based on inferred topics. Each screenshot is displayed with metadata, including participant ID, app name, and inferred topic labels, along with the generated description. Experts assess the relevance of each screenshot to the assigned topic and provide feedback based on the following questions: "How relevant is this screenshot to the assigned topic?"(Slider: Highly Irrelevant, Irrelevant, Slightly Irrelevant, No Answer, Slightly Relevant, Relevant, Highly Relevant), "How accurate is the description for this screenshot?" (Slider: Highly Inaccurate, Inaccurate, Slightly Inaccurate, No Answer, Slightly Accurate, Accurate, Highly Accurate) Qualitative Feedback: "Please provide any additional comments for your rationale for your answer."

**Within-Cluster Content Similarity.** Experts evaluate the internal consistency of clusters by comparing individual screenshots against a larger sample of images from the same cluster. Each evaluation presents a target image alongside a grid of images from the same cluster. Experts assess consistency through the following questions: "How similar do you think the content of Screenshot left is to the group of screenshots shown on the right? " (Slider: Highly Dissimilar, Dissimilar, Slightly Dissimilar, No Answer, Slightly Similar, Similar, Highly Similar) Qualitative Feedback: "Please explain your reasoning. What factors influenced your similarity rating?"

***Survey 3 - Image Retrieval***

In this survey, experts evaluate image retrieval results for various queries such as “cat,” “nature,” and “therapy.” For each query, the top 5 images retrieved by two models—referred to as Model A and Model B—are presented. Experts review images alongside their similarity scores and descriptions and assess their relevance by answering the question: “How relevant is Image X above to the query ‘Y’ (Model: A/B)?” They respond using a seven-point scale ranging from “Highly Irrelevant” to “Highly Relevant” and can provide additional comments explaining their rationale. After reviewing all images for a query, experts are prompted to identify patterns and trends across the models: “What patterns or trends do you notice in the image retrieval results across different models for the same query?” In the final section, experts are invited to provide free-form queries they would like to test in the image retrieval system by responding to the question: “We are developing a live image retrieval system that can retrieve images based on any user query. When it is ready, what queries would you like to test?”

***More Examples of the Media Content Atlas Topics***

Figure 6 below provides a small subset of the topics that are generated by Media Content Atlas. There are topic labels that range from highly specific, such as Personal Financial Assistance for Incarcerated Grandparents, to more general ones like Personal Conversations and Locations. Although rare, some topic labels generated by LLaMA2 contain formatting inconsistencies, such as "Shawn Mendes Music App Screenshot Label: Shawn Mendes Music App Screenshots Show Recent Song Playback with Duration and Artist Information." Additionally, there are overlapping clusters that could potentially be merged later, such as Music Player Interface Screenshots, Music Player Interfaces, and Music Player Screenshots.

The complete list of clusters and the final visualization will be shared publicly for interactive exploration once it is confirmed and approved that no private or personally identifiable information is present.

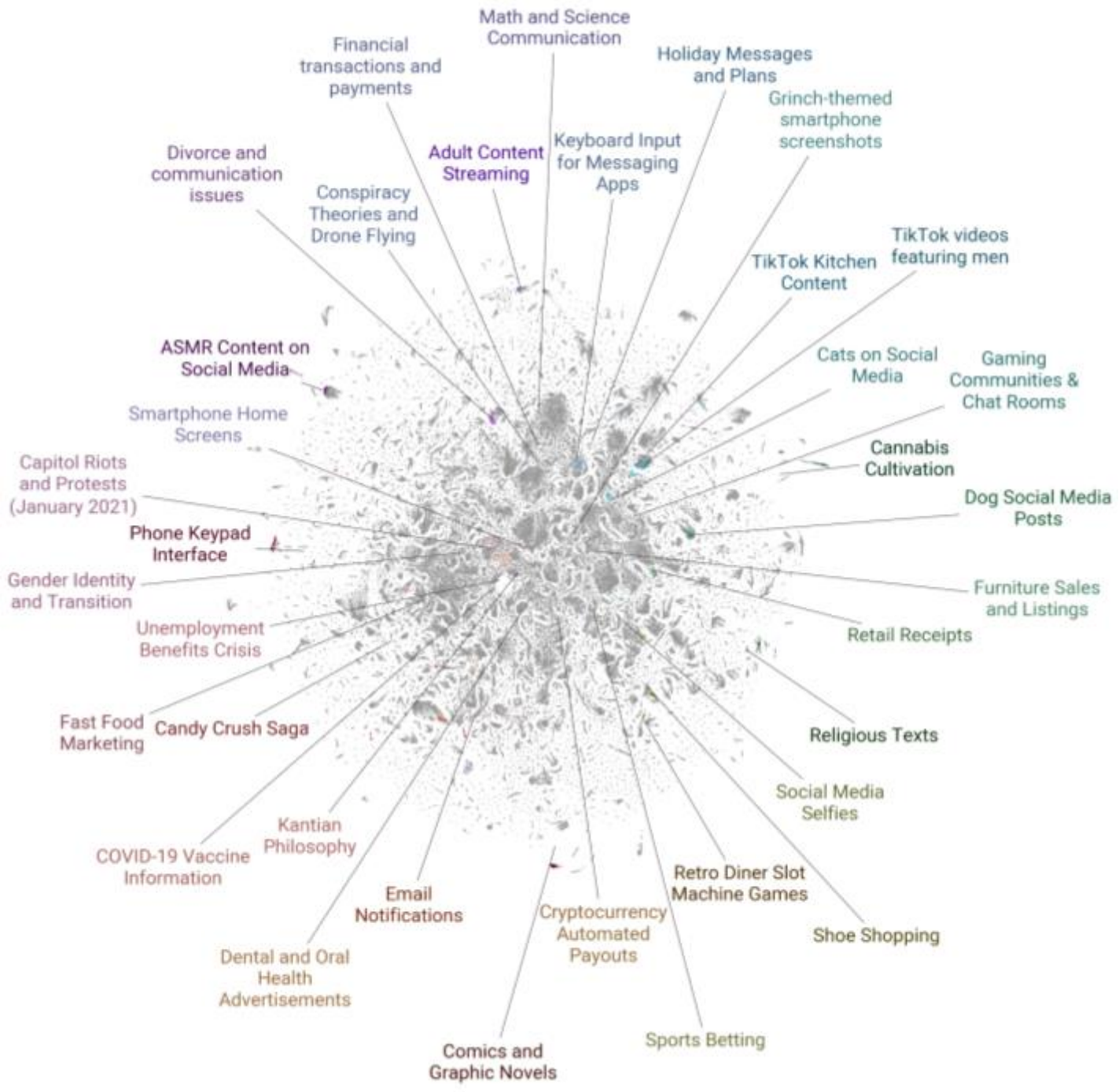

Figure 6: Sample Clusters from Media Content Atlas.

**More Examples of Llava-OneVision Screenshot Descriptions and Expert Comments**

Figure 7 illustrates the accuracy of the generated descriptions across various clusters. The descriptions were generally accurate in capturing key visual elements such as clothing, environments, and objects, as indicated by the green-highlighted areas in the figure. Some descriptions included hallucinated details—contextually plausible elements that were not actually present in the images. For instance, references to steering wheels and dashboards were mentioned in descriptions even when they were not visible. Another notable challenge was perspective misinterpretation; the model described some images as being taken from a first-person viewpoint when they were captured from a third-person perspective. While the descriptions successfully captured key text and UI elements, such as a list of chat contacts with profile pictures and status indicators, they exhibited certain limitations. These included the misinterpretation of icons—such as an incorrectly described red square icon as a heart sign—and missing details like avatars and emojis, which reduced the comprehensiveness of the descriptions. Also, the model exhibited challenges in distinguishing between static images and video calls, often failing to recognize the UI of the apps with live interactions. Some descriptions contained subjective or slightly accurate details, marked in yellow, where the content could be interpreted in multiple ways. For example, the phrase "focusing on the eyes and upper cheek area" was considered slightly accurate because, although the selfie was taken from a slightly upward angle, the entire face was still visible.

***Description 1** - Taken from the cluster "Personalized Image Sharing App"*

The screenshot displays a photo of two women standing in a church, both wearing face masks. The woman on the left is wearing a maroon dress with a floral pattern, while the woman on the right is wearing a white skirt with a black and white patterned top. They are standing in front of a stained glass window and a wooden pew. The text overlay on the image reads "My sweet friend <redacted> I got to see Sunday in <redacted>." The app name is not identifiable.

***Description 2** - Taken from the cluster "Car Eating Scenes"*

The screenshot displays a man sitting in the passenger seat of a car, eating a meal from a white plate (take out box). The man is wearing a black t-shirt and appears to be focused on his food. The car's interior is visible, with the dashboard (not visible) and steering wheel (not visible) in the foreground. The image is taken from the perspective of the driver's seat, looking towards the passenger side (taken from the dashboard side, third-person perspective, looking towards the driver's seat). There is no text present in the image.

***Description 3** - Taken from the cluster "Chat Contact List Organizer"*

The screenshot displays a list of chat contacts on a messaging app. The contacts are named <redacted>. Each contact has a profile picture and a status indicator showing whether they are typing, have a new chat, or have a new snap. The status indicators include a speech bubble for <redacted>, a plus sign(it is an empty arrow-like icon) for <redacted>, a heart (a red square icon) for <redacted>, a checkmark for <redacted>, a heart (a red square icon) for <redacted>, a heart (a red square icon) for <redacted, a heart (a red square icon) for <redacted>, a flower (no icon) for <redacted>, and a camera for <redacted> (camera icon in the lower navigation bar, not for the user). The app name is not identifiable in the screenshot.

***Description 4** - Taken from the cluster "Personalized Image Sharing App"*

The screenshot displays a close-up of a person's face, focusing on the eyes and upper cheek area. The person is wearing glasses with a thin frame, and the lenses reflect a blue light, possibly from a screen. The individual appears to be smiling slightly, and their skin tone is light. There is a caption at the bottom of the image that reads, "LITERALLY A SWEATSHIRT OR LONG TEE WOULD WORK CMO," suggesting a casual or informal context. The app name is not identifiable in this screenshot.

Figure 7: Sample descriptions and expert comments on accuracy. Green highlights indicate highly accurate descriptions, while red font denotes inaccurate details. Corrections to inaccuracies are shown in blue font. Yellow highlights represent descriptions that are slightly accurate or subjective. Participants were not tasked with highlighting or correcting descriptions; this visualization is made for illustration purposes based on observations and participants' responses.

## More examples from Image Retrieval

Despite overall positive ratings from experts, certain queries presented challenges, leading to discrepancies between the models. It is important to note that these observations represent an initial evaluation; a more comprehensive assessment requires calculating recall and precision@k to quantitatively evaluate retrieval performance. The results are influenced not only by the expert ratings and capabilities of the models but also by the availability of relevant screenshots within the repository.

**Blood:** CLIP frequently misidentified ketchup on non-food objects (e.g., shoes) as blood, leading to false positives.

**Coca-Cola:** Llava-OneVision+GTE-Large retrieved general beverage-related images, whereas CLIP identified more Coca-Cola-specific content.

**Influencer:** The ambiguous nature of this query led to inconsistencies, with retrieved images often depicting advertisements or promotional content rather than direct representations of influencers.

**Health:** Llava-OneVision+GTE-Large predominantly retrieved food-related images (e.g., vegetables), whereas CLIP provided more direct health-related content.

**Guns:** Llava-OneVision+GTE-Large returned military-themed imagery, while CLIP better captured direct representations of guns.

**Violence:** Experts interpreted this query as acts of violence, yet CLIP occasionally retrieved unrelated content (e.g., screens with "delete" or "cancel" buttons), whereas Llava-OneVision+GTE-Large provided more relevant violent imagery.

**Substance Abuse:** Both models struggled to distinguish between substance use and abuse, often retrieving images depicting alcohol or cannabis use without clear indicators of abuse.

**Depression:** Llava-OneVision+GTE-Large retrieved more directly relevant depression-related content, whereas CLIP returned images related to symptoms or associated concepts (e.g., anxiety, dark themes), which were deemed less direct.